\documentclass[aip,jmp,amsmath,amssymb,preprint,author-year,author-numerical,floatfix]{revtex4-1}
\usepackage{graphicx}
\graphicspath{{./images/}}
\usepackage{dcolumn}
\usepackage{bm}
\usepackage{mathtools}
\usepackage[caption=false]{subfig}
\usepackage{amssymb}
\expandafter\let\csname equation*\endcsname\relax
\expandafter\let\csname endequation*\endcsname\relax
\usepackage{amsmath}
\usepackage{amsbsy}
\usepackage{booktabs}

\begin{document}

\preprint{AIP/123-QED}

\title{Mermin-Wagner theorem, flexural modes, and degraded carrier mobility in 2D crystals with broken horizontal mirror 
      (${\boldsymbol{\sigma}}_{\rm h}$) symmetry}
\author{Massimo V. Fischetti}
\author{William G. Vandenberghe}
\email{William.Vandenberghe@utdallas.edu.}
\affiliation{Department of Materials Science and Engineering, The University of Texas at Dallas\\
             800 W. Campbell Rd., Richardson, TX 75080}

\date{\today}
\begin{abstract}
We show that the electron mobility in ideal, free-standing two-dimensional `buckled' crystals with broken horizontal mirror ($\sigma_{\rm h}$)
symmetry and Dirac-like dispersion (such as silicene and germanene) is dramatically affected by scattering with the acoustic flexural modes 
(ZA phonons). This is caused both by the broken $\sigma_{\rm h}$ symmetry and by the diverging number of long-wavelength ZA phonons, consistent 
with the Mermin-Wagner theorem. Non-$\sigma_{\rm h}$-symmetric, `gapped' 2D crystals (such as semiconducting transition-metal dichalcogenides with a tetragonal crystal structure) are affected less severely by the broken $\sigma_{\rm h}$ symmetry, but equally seriously 
by the large population of the acoustic flexural modes. We speculate that reasonable long-wavelength cutoffs needed to stabilize the structure 
(finite sample size, grain size, wrinkles, defects) or the anharmonic coupling between flexural and in-plane acoustic modes (shown to be effective in
mirror-symmetric crystals, like free-standing graphene) may not be sufficient to raise the electron mobility to satisfactory values. 
Additional effects (such as clamping and phonon-stiffening by the substrate and/or gate insulator) may be required.
\end{abstract}

\keywords{Two-dimensional materials, flexural modes, Mermin-Wagner theorem, electron mobility, silicene, germanene}

\maketitle

\section{Introduction}

\begin{figure}[tb]
\includegraphics[width=8cm]{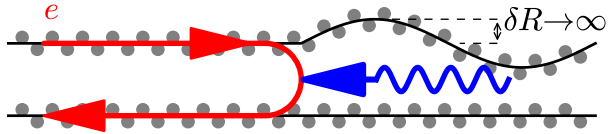}
\caption{Cartoon illustration of the scattering process of an electron absorbing a flexural mode. The upper half of the figure illustrates an 
         incoming electron and a phonon before scattering, the bottom half shows the scattered electron after it has absorbed the phonon 
         (wavy line). Whereas in materials with horizontal mirror ($\sigma_{{\rm h}}$) symmetry, such as graphene or semiconducting TMDs in the hexagonal phase, 
         this scattering process is prohibited by symmetry, in materials without $\sigma_{\rm h}$-symmetry such as silicene or 
semiconducting TMDs in the tetragonal phase, 
         scattering with flexural modes is allowed. This scattering can be exceedingly strong since the out-of-plane ionic displacement diverges as $1/Q$ 
         at a small phonon momentum $Q$ ($\delta R_{{\rm ZA}}(Q) \propto 1/Q\propto \lambda \to \infty$, instead of the `usual' in-plane behavior $\propto 1/Q^{1/2}$)
         and the number of occupied phonon modes, governed by the Bose-Einstein distribution, diverges as well 
         ($N_{{\rm ZA}}(Q) \propto 1/Q^2 \to \infty$).}
\label{fig:scattering}
\end{figure}

Two-dimensional materials -- such as graphene\cite{Geim_2007}, silicene\cite{Vogt_2012,Tao_2015,Houssa_2011}, 
germanene\cite{Houssa_2011,Davila_2014}, phosphorene\cite{Liu_2014,Gomez_2014,Li_2014}, 
stanene\cite{Xu_2013,Zhu_2015,Vandenberghe_2014,Vandenberghe_2014a}, and  (semiconducting)
transition-metal dichalcogenides (TMDs)\cite{Mak_2010}, just to mention the most `popular' in a growing list -- have stimulated interest and excitement. On scientific grounds, most of these materials exhibit novel topological properties linked, for example, to the Quantum Spin Hall 
effect\cite{Haldane_1988,Kane_2005,Bernevig_2006} and superconductivity\cite{Bernevig_2013}. From a more practical perspective, they promise to open 
the door to continued scaling of `conventional' field-effect transistors (FETs)\cite{Fischetti_2013} and to a plethora of other nanoelectronic and
optoelectronic applications. From this perspective, most of the promise offered by these materials obviously hinges on a `satisfactory' mobility 
of the charge carriers, `satisfactory' meaning {\em at least} several hundreds cm$^{2}$/Vs to be competitive with Si in digital logic applications. 
Graphene satisfies and even largely exceeds this requirement. Other materials, instead, so far have shown an electron 
mobility that is, at best, no more than satisfactory. In many cases, this is simply the result of yet immature material technology: 
Defects, grain boundaries, charge impurities, interactions with the supporting substrate or gate insulators, all of these may be blamed. 
Here, however, we argue that there may be more to it. Indeed, our main goal is to point out the existence of an {\em intrinsic} physical problem that 
arises theoretically: This problem is strictly linked to the thermodynamic instability of ideal two-dimensional crystals and it affects 
the electron mobility in materials in which electrons  can be scattered by acoustic flexural modes via one-phonon processes, 
as illustrated schematically in Fig.~\ref{fig:scattering}. Such scattering is possible in two-dimensional crystals that are not symmetric under
reflections on the horizontal plane of the lattice. We shall refer to these crystals as being non-$\sigma_{{\rm h}}$-symmetric.

We should emphasize immediately that this problem has been already discussed at length in the past in the context of biological membranes, polymerized layers,
and some inorganic surfaces\cite{Nelson_1987,Nelson_1989}. The problem has obviously attracted renewed attention with the advent of 
graphene\cite{Kownaki_2009,Gazit_2009,Gazit_2009a,Braghin_2010,Lebedev_2012,Gornyi_2015}. In particular, recent work has solved the problem
of the diverging electron/two-phonon scattering processes in $\sigma_{{\rm h}}$-symmetric crystals\cite{Gornyi_2012}, explaining the large electron mobility observed in free-standing graphene\cite{Bolotin_2008,Castro_2010}. This `state-of-the-art' will be mentioned again below. 
Here, we discuss the serious additional problem that arises exclusively in 'non-graphene-like' systems, namely, in non-$\sigma_{{\rm h}}$-symmetric 
two-dimensional crystals.

Since we will be forced to deal with divergences, our use of perturbation theory can only give qualitative hints, but compelling hints
nevertheless. Therefore, we will only be able to speculate about possible solutions, intentional or `natural', of the problem, assuming they exist. 
Our discussion will be restricted to two-dimensional lattices of the honeycomb type, {\em i.e.} hexagonal and tetragonal crystals.

We consider here two main issues with two-dimensional crystals: The parabolic nature of the dispersion of acoustic out-of-plane (flexural)
modes (ZA phonons) and the degeneracy of the bands at the K/K$^\prime$ symmetry point in hexagonal/tetragonal crystals with a Dirac-like electron dispersion.
Consequences of the former issue are related to the well-known `stability problem', as we have mentioned above. On the contrary, to our knowledge 
the effect of the latter in non-$\sigma_{\rm h}$-symmetric crystals has received so far little or no attention.\\ 

{\em 1. Parabolic dispersion of the ZA phonons.} The first issue originates from the well-known -- and a bit controversial -- 
theorem known as the {\it Mermin-Wagner-Hohenberg-Coleman theorem}\cite{Mermin_1966,Hohenberg_1967,Coleman_1973}, although similar arguments had
been raised earlier by Peierls\cite{Peierls_1935} and Landau\cite{Landau_1937}. We shall simply
refer to this theorem as the `Mermin-Wagner' theorem, for brevity. This theorem, which we shall briefly revisit below, states that 
{\em infinite, ideal} two-dimensional (2D) crystals are not stable: The thermal population of long-wavelength out-of-plane (or `flexural') acoustic 
phonons diverges at any finite temperature. The theorem does not go as far as providing a quantitative estimate for the maximum size 
we may hope to obtain for a stable 2D crystal. But, probably, the `wrinkles', defects, and finite grain size that we normally associate with samples of 2D crystals, are direct consequences of the Mermin-Wagner theorem, as also suggested by Geim and Novoselov\cite{Geim_2007}. 
It may seem at first that this problem may affect the stability of the crystals, but that it should be unrelated to the carrier mobility. 
On the contrary, it is strongly connected to
the electron-transport properties of non-$\sigma_{\rm h}$-symmetric systems in which electrons can interact with flexural phonons.
Indeed, the reason for the instability of ideal, infinite 2D crystals ultimately stems from the parabolic dispersion of the acoustic 
flexural modes, the ZA phonons. This is an unavoidable consequence of their 2D nature and it is the reason why their population diverges 
at long wavelength. 

To see how in non-$\sigma_{\rm h}$-symmetric crystals 
the transport properties are affected by this, consider a 2D crystal, an electron of wavevector ${\bm K}$, 
and calculate the rate at which it emits or absorbs an acoustic phonon using the usual elastic\cite{elastic}, high-temperature equipartition 
approximation\cite{Jacoboni_1983}:
\begin{equation}
\frac{1}{\tau({\bm K})}  = \frac{\Delta_{\rm ac}^{2} k_{\rm B}T}{2 (2\pi)^{2} \hbar \rho} \
    \int_{0}^{2 \pi} {\rm d} \phi \ \int_{0}^{\infty} {\rm d} Q \ \frac{Q^{3}}{\omega_{\bm Q}^{2}} \ 
                                                              \delta [ E_{\bm K} - E_{{\bm K}+{\bm Q}} ] \ ,  
\label{eq:intro}
\end{equation}
where $\rho$ is the areal mass density of the crystal, $\Delta_{\rm ac} Q$ is the acoustic `deformation potential' expressing the strength of the 
electron-phonon interaction, $E_{\bm K}$ is the electron dispersion, ${\bm Q}$ is the phonon wavevector, and, finally, $\omega_{\bm Q}$ is
the phonon dispersion. (We use upper-case symbols for 2D vectors.)
It is immediately clear that the `usual' linear phonon dispersion involving the sound velocity $c_{\rm s}$, 
$\omega_{\bm Q} \sim c_{\rm s} Q$ can be integrated straightforwardly. On the contrary, a parabolic dispersion $\omega_{\bm Q} \sim b \ Q^{2}$
results in a logarithmic singularity at small scattering angles $\phi$, since the integrand behaves as $1/Q$. This divergence is directly
connected to the Mermin-Wagner theorem. Physically, the diverging scattering rate represents an electron interacting with a diverging 
number of ZA phonons.

Two-dimensional crystals with a horizontal mirror plane (that we shall
assume is the plane $(x,y)$, the $h$-plane) are immune to this problem to first order (in the atomic displacement):
As a consequence of their $\sigma_{{\rm h}}$ symmetry, the ionic
displacement associated with a ZA phonon is always perpendicular to
the $h$-plane and the potential associated with the flexural displacement
is odd with respect to $\sigma_{{\rm h}}$. Moreover, because each
band is either even or odd with respect to $\sigma_{{\rm h}}$, the
product of the initial state, the potential due to the displacement
and the final state is odd with respect to $\sigma_{{\rm h}}$. The
electron-phonon matrix element, being the integral of an odd function,
thus vanishes and intraband electronic transitions assisted by flexural
modes (ZA phonons or their optical counterpart, ZO) are forbidden.
Graphene and semiconducting TMDs with horizontal mirror symmetry (namely, those with the hexagonal crystal structure) are thus spared from intra-band ZA phonon scattering to first order. The 300~K electron mobility in these symmetric systems (such as free-standing graphene) may be limited to a 
few $10^4\,{\rm cm}^2/({\rm Vs})$ by two-phonon processes\cite{Castro_2010,Mariani_2008,Mariani_2010} (still quite a large value, because the interaction is second-order in the ionic displacement, as symmetry demands the simultaneous emission or absorption of two phonons). However, even
these processes have been shown to be strongly weakened by the anharmonic coupling of the flexural modes to the in-plane modes. Indeed, in these symmetric crystals, this coupling stiffens the crystal\cite{Gornyi_2015} and removes the electron/ZA-phonon singularity at the Dirac point\cite{Gornyi_2012}. However, in crystals without a horizontal mirror plane, such as silicene,
germanene, or semiconducting TMDs with a tetragonal (T) or distorted-tetragonal (T') crystal structure, this protection
against first-order, one-phonon scattering processes by flexural modes 
is absent and, as we shall briefly discuss below, the anharmonic coupling of the flexural modes to the
in-plane modes does not eliminate the singularity and it is largely ineffective in boosting the mobility that is now controlled 
by one-phonon processes. These non-$\sigma_{\rm h}$ materials are therefore much more seriously affected by the diverging number of 
long-wavelength ZA phonons.\\

{\em 2. Dirac-like dispersion.} 
The second very important issue we consider is a problem that affects non-$\sigma_{\rm h}$-symmetric materials with a Dirac-like electron dispersion: In these materials, we must face the additional problem of an increased strength of the electron/ZA-phonon coupling. This results
from the degeneracy of the bands at the symmetry-point K, where back-scattering with flexural modes is orders of magnitudes stronger 
compared to scattering with other modes, as discussed in the next section. 
Unless some long-wavelength cutoff is provided by other mechanisms (damping of the ZA modes by a substrate and/or gate insulator, finite and
small grain size, defects, wrinkles, and such), the expected mobility will be zero `for all practical purposes'.\\

We are aware only of a few previous calculations of the electron mobility in silicene\cite{Li_2013,Shao_2013}, 
germanene\cite{Li_2013,Ye_2014} and transition-metal dichalcogenides with a tetragonal crystal structure ({\em i.e.}, HfSe$_{2}$, HfS$_{2}$,  
ZrSe$_{2}$, and ZrS$_{2}$ in Ref.~\onlinecite{Zhang_2014}). 
In Ref.~\onlinecite{Li_2013}, the divergent scattering rates were artificially `regularized' by 
approximating the parabolic dispersion of the ZA modes with a linear dispersion. Even with these simplifications, Li {\em et al.}\cite{Li_2013}
found the ZA phonons to dominate the mobility, but the 300~K values reported are orders of magnitude larger than what we find here, thanks to 
the {\em ad hoc} regularization of the singularity. Shao {\em et al.}\cite{Shao_2013} and  
Ye and co-workers\cite{Ye_2014} have calculated the electron mobility in silicene and germanene, respectively, using an approximate 
expression that involves only the 2D in-plane elastic constant and deformation potentials. This amounts to neglecting the effect of the
flexural modes and of the Mermin-Wagner theorem altogether. Moreover, even when dealing with the interaction
of electrons with in-plane phonons, only longitudinal acoustic phonons are considered (via the use of the $c_{11}$
elastic constant) and wavefunction-overlap effects are ignored. 
In so doing, they have obtained an electron mobility at 300~K of $2.5 \times 10^{5}$ cm$^{2}$/Vs 
for silicene, of $6 \times 10^{5}$ cm$^{2}$/Vs for germanene. These values approach and even exceed the calculated mobility in graphene. Experimental values for silicene\cite{Tao_2015} are far less exciting, showing a field-effect electron mobility of about 
80-to-100 cm$^{2}$/Vs. 
Note that in these cases the 2D crystal was supported. The 
acoustic-phonon-limited electron mobility at 300~K in HfSe$_{2}$, HfS$_{2}$, ZrSe$_{2}$, and ZrS$_{2}$ has been calculated to be as high as 
3579, 1833, 2316, and 1247 cm$^{2}$/Vs, respectively, in Ref.~\onlinecite{Zhang_2014}. However, these results have been obtained using the same 
model employed by Shao {\em et al.}\cite{Shao_2013} and Ye and co-workers\cite{Ye_2014}, therefore ignoring in-plane transverse modes,  
wavefunction-overlap effects and, most important, the coupling of electrons with acoustic flexural modes in these 2D crystals with a 
non-$\sigma_{\rm h}$-symmetric tetragonal structure. 
Finally, values ranging from 286 cm$^2$/(Vs) (hole mobility in FETs\cite{Liu_2014}) to about 900 cm$^2$/(Vs) (electron mobility in quantum-well FETs\cite{Tayari_2015}) have been reported for the (gapped) phosphorene. However, whereas this crystal is not strictly $\sigma_{\rm h}$ symmetric, its space group is non-symmorphic and contains a glide plane with horizontal mirror symmetry. Therefore, its point group contains the $\sigma_{\rm h}$ symmetry and the ZA/electron coupling is forbidden, much like in the case of graphene. All puckered 2D crystals of this structure are therefore immune from this problem.
 Much more relevant to our study is the recent {\em ab initio} work by Gunst {\em et al.}\cite{Gunst_2015}:
In this work, the electron/ZA-phonon interaction in silicene has been fully characterized, but it has been omitted when computing the electron mobility 
on the basis of the difficulty of accounting correctly for dielectric screening and speculations on possible effect of the substrate on the dispersion of the ZA phonons. We shall revisit the former issue in a footnote, the latter at length in the following.

Flexural modes have also been theoretically found to affect very strongly thermal transport in graphene\cite{Lindsay_2010,Singh_2011}
(although somewhat controversially\cite{Nika_2009,Kong_2009}) and electronic transport in thin, free-standing quantum wells by Bannov {\em et al.}\cite{Bannov_1995} and
by Glavin and co-workers\cite{Glavin_2002}. Indeed, in these studies, the momentum and energy electron relaxation rates have been found to be 
enhanced in free-standing slabs because of the parabolic dispersion of the acoustic flexural modes, while supported films exhibit exponentially damped
out-of-plane modes. Their conclusions apply only at temperatures lower that $T^{\ast} = 2 \pi \hbar c_{\rm s}/(k_{\rm B} W)$, 
where $W$ is the thickness of the quantum well. However, for the 2D crystals we consider here, this temperature diverges, since phonon `confinement' along the out-of-plane direction is an ill-defined concept and we should consider the limit $W \rightarrow 0$. 

To estimate the value of the mobility, we compute the deformation
potentials from first principles as described in Ref.~\onlinecite{Vandenberghe_2015}.
We use the Vienna Ab Initio Software Package (VASP) \cite{VASP1,VASP2,VASP3,VASP4}
to compute the wavefunction of the initial and final state and perform
an inverse Fourier transform to get the wavefunction in real space.
We obtain the change in potential associated with a phonon displacement
by calculating the potential for a small displacement of each atom along each Cartesian direction
and subsequently multiplying it by the polarization
vector. Phonon energies and polarization vectors are calculated using a post-processing
package\cite{Togo2015}.  The result of this calculation is a deformation potential,
$DK$, whereas the total electron-phonon interaction requires the multiplication
with a factor measuring the ionic displacement $\sqrt{\hbar/(2\rho\omega)}$ of a mode of frequency $\omega$,
as well as the Bose-Einstein distribution $N(E)=(e^{E/(kT)}-1)^{-1}$.\cite{screening}

We proceed as follows: We consider first the electron/ZA-phonon coupling in 2D crystals. We then review the Mermin-Wagner theorem and  
provide quantitative estimates for the room-temperature electron mobility in silicene. We conclude by speculating on the
nature of a long-wavelength cutoff that may result in an increased electron mobility. Our main conclusion can be summarized as follows:
ideal, free-standing 2D crystals with a broken $\sigma_{\rm h}$ symmetry should exhibit an extremely low electron mobility. Deviations from this
conclusion would suggest the existence of some yet-to-be-determined mechanism able to damp the acoustic flexural modes. 

\section{Electron coupling to flexural modes in Dirac materials}

To investigate the nature of the interaction with flexural modes in
materials with a Dirac cone, we first resort to a simple tight-binding
model and then proceed by applying the rigid-ion approximation\cite{Ziman_1958}.\\

{\em 1. Tight binding model.} In the simplest tight-binding model, the basis consists of two orbitals
with their respective origin at each of the two atoms A and B in each
unit cell. For the wavefunctions composed of these orbitals to be
compatible with the crystal symmetry, the orbitals must have a rotation
axis perpendicular to the $h$-plane and the orbitals at A and B must
be each others' inversion image. In the absence of a $\sigma_{\rm h}$-symmetry,
the orbitals need not have a horizontal mirror symmetry and will resemble
an sp$^{3}$ orbital rather than a p$_{z}$ orbital. The electronic
tight-binding Hamiltonian has the same form as for graphene,
$H=\hbar \upsilon_{\rm F}\left(K_{x}\sigma_{x}+K_{y}\sigma_{y}\right)$
(where $\boldsymbol {\sigma}$'s are the Pauli matrices, $\upsilon_{\rm F}$ is the Fermi velocity,
and the 2D wavevector ${\bf K}$ is measured from the K symmetry point)
and has eigenvalues $E=\hbar \upsilon_{{\rm F}}\sqrt{K_{x}^{2}+K_{y}^{2}}$
and pseudospin eigenvectors $[1,{\rm e}^{{\rm i}\phi}]/\sqrt{2}$. 

However, considering a rigid out-of-plane displacement of the potential
$\delta V$ associated with long-wavelength flexural modes, a non-vanishing
matrix element $V_{0}=\langle A|\delta V|A\rangle$ is obtained. (In
the notation used below, $V_{0}^{2}=1/(2\Omega)(DK)^{2}\hbar/(2\rho\omega_{\bm{Q}}^{{\rm (ZA)}})(N_{\bm{Q}}^{{\rm (ZA)}}+1/2\pm1/2)$
for emission (+) and absorption (-) processes.) Because of inversion
symmetry, $\langle A|\delta V|A\rangle=-\langle B|\delta V|B\rangle$.
Therefore, in the tight-binding basis, the electron-phonon Hamiltonian
for the ZA phonons, $\widehat{H}_{{\rm ep}}$, is diagonal with elements
$H_{{\rm ep},11}=V_{0}$ and $H_{{\rm ep},22}=-V_{0}$. As a result,
the matrix element between an initial state $\mid1\rangle=[1,e^{{\rm i}\phi_{1}}]/\sqrt{2}$
and a final state $\mid2\rangle=[1,e^{{\rm i}\phi_{2}}]/\sqrt{2}$
is $M_{12}=\langle2\mid H_{{\rm ep}}\mid1\rangle=V_{0}\left(1-e^{{\rm i}(\phi_{1}-\phi_{2})}\right)/2$
whose magnitude is $|M_{12}|=V_{0}|{\rm sin}((\phi_{1}-\phi_{2})/2)|$.
Now, the atomic displacement, and consequently the magnitude of the potential
$|\delta V|$, increases as $\omega_{{\bf Q}}$ decreases. So when
the initial and final states approach each other at the K symmetry-point in reciprocal space,
${\bf Q}$ and $\omega_{{\bf Q}}$ decrease and the coupling of the
electron-phonon interaction diverges. This is contrary to the `usual' acoustic phonon
scattering where a decrease of the deformation potential $DK=\Delta Q$
counteracts the increase in displacement and the increase in occupation
due to the Bose-Einstein distribution. The result is that the electron-phonon
matrix element $M\propto DK^{2}/\omega_{{\bf Q}}^{2}$ remains constant
as the initial and final state approach each other.

The absence of a vanishing interaction can also be understood from
a group-theoretical point of view. First of all, time-reversal symmetry
combined with inversion symmetry map each point in the Brillouin zone
onto itself. In the absence of spin-orbit coupling and degeneracy, this ensures that
the periodic part of the Bloch waves $u^{\ast}_{\bf K}({\bf r}) \propto u_{\bf K}(-{\bf r})$. In addition, acoustic displacements are odd under inversion and therefore, the electron-phonon matrix elements $\langle 1 \mid H_{\rm ep} \mid 1 \rangle$, being the integral of a function which is odd with respect to inversion,
vanish at each point in the Brillouin zone where bands are non-degenerate,
but are nonzero at the K-point, where bands \textit{are} degenerate. Furthermore,
the K-point does not have inversion symmetry but has trivial symmetry
(E), a 3-fold rotation symmetry (2C$_{3}$) and 3 vertical mirror
planes (3$\sigma_{\rm v}$) which are described by the $D_{3}$ point
group. $D_{3}$ is isomorphic to the dihedral group which is the smallest
non-Abelian group and therefore also the smallest group which has
a two-dimensional irreducible representation (E), needed for the
Dirac cone, next to the trivial representation ($A_{1}$) and an odd
representation ($A_{2}$). The out-of-plane displacements associated
with the flexural modes transform onto themselves under all 6 symmetry
operations of $D_{3}$, they are represented by the trivial representation
$A_{1}$ which offers no symmetry protection. 

Therefore the interaction of electrons with flexural
phonons at the Dirac-cone is not prohibited by the time-reversal/inversion
symmetry, because the states are degenerate; nor it is prohibited by the symmetry
of the out-of-plane displacements, since the latter are represented
by the trivial representation. This results in strong electron-phonon scattering and so in a reduced carrier mobility\\

{\em 2. Rigid-ion approximation.} In the rigid-ion approximation, the matrix element for the electron-phonon Hamiltonian, 
${\widehat{H}}_{\mathrm{ep}}^{(\eta)}$, for a phonon of branch $\eta$ 
(acoustic, optical; longitudinal, transverse, or out-of-plane/flexural) and wavevector ${\bm q} = ({\bm Q},0)$ 
between initial and final Bloch states with wavevectors ${\bm K}$ and ${\bm K}'$ in bands $n$ and $n'$, respectively, can be written 
as\cite{Fischetti_2013a}:    
\begin{equation}
\langle {\bm K}' n' | {\widehat{H}}_{\mathrm{ep}}^{(\eta)} | {\bm K} n \rangle \ = \
  \sum_{{\bm G}, {\bm G}'} \ {\mathcal{D}}^{(\eta)}_{{\bm K}'-{\bm K},{\bm G}} \
       u^{(n')*}_{{\bm K}', {\bm G}' + {\bm G}} \  u^{(n)}_{{\bm K}, {\bm G}'} \ ,
\label{eq:Hep}
\end{equation}
where the `coupling constant' ${\mathcal{D}}^{(\eta)}_{{\bm Q},{\bm G}}$ is: 
\begin{equation}
{\mathcal{D}}^{(\eta)}_{{\bm Q},{\bm G}} \ = \ i ( {\bm q}+{\bm G} ) \ \cdot \ 
   \sum_{\gamma} \ {\bm e}^{(\gamma)}_{{\bm Q},\eta} \ V^{(\gamma)}_{{\bm q }+{\bm G}} \
         \mathrm{e}^{ i ({\bm q} + {\bm G}) \cdot {\mbox{\boldmath{$\tau$}}}_{\gamma}} \ {\mathcal{A}}_{{\bm Q},\eta} \ .
\label{eq:Dep}
\end{equation}
The functions $u^{(n)}_{{\bm K}, {\bm G}}$ are Fourier components of the periodic part of the Bloch waves, the 
vectors ${\bm G}$ are the reciprocal-lattice vectors, 
$V^{(\gamma)}_{\bm q}$ is the (pseudo)potential of ion $\gamma$ in the unit cell, 
and ${\bm e}^{(\gamma)}_{{\bm Q}, \eta}$ is the unit displacement of ion $\gamma$ caused by a phonon with wavevector ${\bm Q}$ in branch $\eta$.
Note that we have explicitly isolated the very important phase-factor $\exp({\mathrm i} {\mathbf q} \cdot {\mbox{\boldmath{$\tau$}}}_{\gamma})$ 
from the phonon polarization vector ${\bm e}^{(\gamma)}_{{\bm Q},\eta}$. In Eq.~(\ref{eq:Hep}), the quantity
${\mathcal{A}}_{{\bm Q},\eta}$ represents the ionic displacement and is given by:
\begin{equation}
{\mathcal{A}}_{{\bm Q},\eta}^{2} \ = \ 
  \left ( \frac{\hbar}{2 \rho \ \omega_{{\bm Q},\eta}} \right ) \
    \left \{ \begin{array}{c}
              N_{{\bm Q},\eta} \\
              1 + N_{{\bm Q},\eta} 
             \end{array} \right \} \ . 
\label{eq:Aep}
\end{equation}
In this expression, $N_{{\bm Q},\eta}$ is the occupation number of phonons of wave vector ${\bm Q}$ and branch $\eta$ and the upper (lower)
symbol within curly brackets should be taken for absorption (emission) processes. 

Consider the simple but relevant case of a crystal with two identical atoms in each unit cell ({\em e.g.}, graphene, silicene, germanene, stanene). The origin of the coordinate system can be taken mid-way between the two atoms, so their positions will be 
${\mbox{\boldmath{$\tau$}}}_{1} = - {\mbox{\boldmath{$\tau$}}}_{2} = {\mbox{\boldmath{$\tau$}}}$. 
For intraband transitions, in the limit $Q \rightarrow 0$ (${\bm K} \rightarrow {\bm K}'$), the phonons at $\Gamma$ will have 
in-phase (acoustic) or out-of-phase (optical) polarizations, that is,  
${\bm e}_{\eta} = {\bm e}^{(1)}_{{\bm Q},\eta} = \pm {\bm e}^{(2)}_{{\bm Q},\eta}$, the plus (minus) 
sign applying to acoustic (optical) branches. In the sum over ${\bm G}$ in Eq.~(\ref{eq:Hep}), for every vector 
${\bm G}$ we can consider the sum of each pair of terms corresponding to ${\bm G}$ and $-{\bm G}$. 
Then, to anticipate the obvious conclusion, after summing over the ions in the unit cell,
each pair of $\pm {\bm G}$-vectors will result in a contribution of the form 
$\sim {\bm e} \cdot {\bm G} \ \{ \cos [ ({\bm q} + {\bm G}) \cdot \ {\mbox{\boldmath{$\tau$}}}] -
                         \cos[ ({\bm q} - {\bm G}) \cdot \ {\mbox{\boldmath{$\tau$}}}] \}$ (that tends to zero as $Q \rightarrow 0$) for the acoustic branches, and a contribution
$\sim {\bm e} \cdot {\bm G} \ \{ \sin [ ({\bm q} + {\bm G}) \cdot \ {\mbox{\boldmath{$\tau$}}}] -
                         \sin [({\bm q} - {\bm G}) \cdot \ {\mbox{\boldmath{$\tau$}}}] \}$ (that tends to a constant as $Q \rightarrow 0$) for the  optical branches. More specifically, each pair will give a contribution
\begin{align}
2 \ {\mathrm i} \ V_{\bm G} \ {\mathcal{A}}_{{\bm Q},\eta} \  {\bm e}_{\eta} \cdot 
        & [ {\bm q} \ \cos ( {\bm G} \cdot {\mbox{\boldmath{$\tau$}}} ) \ 
      + \ {\bm G} \ ({\bm q} \cdot {\mbox{\boldmath{$\tau$}}}) \ ] \nonumber \\  
&              \times \ \sum_{{\bm G}'} u^{(n)*}_{{\bm K}, {\bm G}'+{\bm G}} \ u^{(n)}_{{\bm K}, {\bm G}'}  \
\label{eq:acoustic}
\end{align} 
for acoustic phonons, and a contribution
\begin{align}
2 \ V_{\bm G} \ {\mathcal{A}}_{{\bm Q},\eta} \ & \sin ( {\bm G} \cdot {\mbox{\boldmath{$\tau$}}}) \
         ({\bm q} + {\bm G}) \cdot {\bm e}_{\eta} \nonumber \\ 
&              \times \ \sum_{{\bm G}'} u^{(n)*}_{{\bm K}, {\bm G}'+{\bm G}} \ u^{(n)}_{{\bm K}, {\bm G}'}  \
\label{eq:optical}
\end{align}
for optical phonons. Note that, in order to obtain these expressions, we have made use of the symmetry property 
\begin{equation}
\sum_{{\bm G}'} u^{(n)*}_{{\bm K}, {\bm G}'+{\bm G}} \ u^{(n)}_{{\bm K}, {\bm G}'} \  = \
\sum_{{\bm G}'} u^{(n)*}_{{\bm K}, {\bm G}'-{\bm G}} \ u^{(n)}_{{\bm K}, {\bm G}'} \ \ . 
\label{eq:symmetry}
\end{equation}
This reflects the lattice symmetry on the plane of the layer and holds true for all in-plane ${\bm G}$-vectors (that is, with $G_{z} = 0$). 
When considering in-plane modes, these are the only terms one has to consider inside the sum over ${\bm G}$-vectors in Eq.~(\ref{eq:Hep}). 
Equations~(\ref{eq:acoustic}) and (\ref{eq:optical}) correspond to the well-known fact that at small $Q$, the acoustic `deformation 
potential', $DK$ (the quantity given by Eq.~(\ref{eq:Hep}), but without the 'ionic displacement' factor ${\mathcal{A}}_{{\bm Q},\eta}$)  
vanishes linearly with $Q$ ($DK \approx \Delta_{\mathrm{ac}} Q$), whereas the optical deformation potential approaches a constant, usually denoted by $DK_{0}$.

Considering now the out-of-plane (flexural) modes ZA and ZO, the situation is different, depending on whether we consider 2D crystals
that are $\sigma_{\rm h}$-symmetric, or crystals for which this symmetry is broken.

For $\sigma_{\rm h}$-symmetric systems, such as graphene, the flexural modes have purely out-of-plane polarization vectors 
${\bm e}_{\mathrm{ZA,ZO}}$ and $u^{(n)}_{{\bm K},{\bm G}} = u^{(n)}_{{\bm K},{\bm G}'}$ for ${\bm G}' = (G_{x},G_{y},-G_{z})$. Therefore, 
every pair of such vectors inside the sum over ${\bm G}$ in Eq.~(\ref{eq:Hep}) will give a vanishing contribution. All intraband processes 
have a vanishing matrix element.
Electrons and flexural modes are decoupled at first order.

On the other hand, for buckled 2D crystals in which the $\sigma_{\rm h}$ symmetry is broken, the lack of this symmetry results in a lack of cancellation between each pair of ${\bm G}$ and $-{\bm G}$ vectors, giving rise to a coupling between electrons and out-of-plane modes. 
Moreover, as we have seen using the tight-binding model, for layers with a Dirac-like dispersion, the deformation potential is strongly enhanced. 
At the K symmetry point, since the bands are degenerate, regardless of the basis chosen in this eigen-subspace, the lack of
$\sigma_{\rm h}$ symmetry does not yield the cancellation of terms that results in a matrix element vanishing with $Q$, as shown in 
Eq.~(\ref{eq:acoustic}). Therefore, the acoustic deformation potential, $DK$, becomes a 
$Q$-independent quantity as we approach the symmetry point K and must take a form of the type
$\Delta (E)Q$ with $\Delta (E) \sim 1/E $, where $E = \hbar \upsilon_{\mathrm {F}} K$ is the electron dispersion in proximity of the K-point 
expressed in terms of the Fermi velocity $\upsilon_{\mathrm {F}}$ and measuring the electron wavevector ${\bm K}$ 
from the K-point. 

Indeed, tight-binding arguments we have already presented and
also density functional theory (DFT) calculations we have performed, as performed
in Ref.~\onlinecite{Vandenberghe_2015} and briefly discussed in the introduction,
show that for electronic initial and final states ${\bm{K}}$ and
${\bm{K}}'$ on the energy-conserving shell $K=K'$, the acoustic
deformation potentials can be approximated by 
\begin{align}
DK_{{\rm ZA}}\  & \sim\ (DK)_{0}\ \sin(\phi/2)\ \ \ \ \mbox{(intra-band)}\nonumber \\
DK_{{\rm ZA}}\  & \sim\ (DK)_{0}\ \cos(\phi/2)\ \ \ \ \mbox{(inter-band)}\ ,\label{eq:DKac_1}
\end{align}
where $\phi$ is the scattering angle between ${\bm{K}}$ and ${\bm{K}}'$.
Note that this expression is valid \textit{regardless of the electron
energy}. This has a strong implication: Since for small $Q$ we have
$K\approx Q\phi$, Eq.~(\ref{eq:DKac_1}) for intra-band transitions can be rewritten as: 
\begin{equation}
DK_{{\rm ZA}}\ \sim\frac{(DK)_{0}\hbar\upsilon_{\mathrm{F}}}{2E}\ Q\ .\label{eq:DKac_2}
\end{equation}
Equations~(\ref{eq:DKac_1}) and (\ref{eq:DKac_2}) result in a picture
quite different from what holds for scattering with in-plane modes:
Backward scattering is now largely favored, and the strength of the
interaction grows as we approach the K symmetry point.

Therefore, non $\sigma_{\rm h}$-symmetric materials with an energy gap -- and, so, with a parabolic electron dispersion -- 
are not subject to the large coupling with the ZA modes, since
the bottom of the conduction band is non-degenerate. However, the electron mobility will still be affected by the problem mentioned in
the introduction and discussed in the following section: The diverging thermal population of acoustic flexural modes.

\section{The Mermin-Wagner theorem}

In the previous section, we considered the behavior of the deformation
potential due to flexural modes in Dirac materials but, as discussed
in the introduction, scattering rates diverge in all two-dimensional
materials, behavior that is related to the Mermin-Wagner theorem.

A rigorous formulation of this theorem has been given by Coleman in a field-theoretical context\cite{Coleman_1973}: 
{\it Spontaneous breaking of a continuous symmetry is forbidden in $d \le$ 2 dimensions, because the correlation of the corresponding (Goldstone) bosons would diverge.}
In our context, this implies that breaking the continuous translational and rotational symmetries of a homogeneous 2D system required to form
a crystal (that possesses only discrete translational and rotational point-group symmetries), results in the divergence of the thermal
population of ZA phonons. These behave like massive particles, since their dispersion is parabolic. 
This is easily seen with a simple argument: The number of ZA-phonons thermally excited per unit area $\Omega$ is given by: 
\begin{align}
& \frac{\langle N^{\mathrm {(ZA)}} \rangle_{\mathrm {th}}}{\Omega} \  = \
    \int_{BZ} \ \frac{{\mathrm d} {\bm Q}}{(2 \pi)^{2}} \ \left \langle N_{\bm Q}^{\mathrm {(ZA)}} \right \rangle_{\rm th} \nonumber \\
& \approx  \int_{0}^{Q_{\rm BZ}} \frac{{\rm d}Q}{2 \pi} \ Q \ \frac{k_{\rm B}T}{\hbar \omega^{\mathrm{(ZA)}}_{Q}} 
                                      = \frac{k_{\mathrm B}T}{2 \pi \hbar b}  \int_{0}^{Q_{\mathrm {BZ}}} \frac{{\rm d} Q}{Q} \ ,
\label{NZA-thermal}
\end{align} 
an expression that diverges logarithmically as $Q \rightarrow 0$. In this expression we have assumed a parabolic
dispersion for the ZA phonons of the form $\omega^{\rm (ZA)}_{Q} \approx b \ Q^{2}$ and we have approximated the integration over
the Brillouin zone by setting the upper integration limit to some `average' zone-edge wavevector $Q_{\rm BZ}$.
We have also assumed the high-temperature limit, $k_{\rm B}T >> \hbar \omega^{\rm (ZA)}_{\bm Q}$ for all ${\bm Q}$ in the Brillouin zone. 

Comparing Eq.~(\ref{NZA-thermal}) with what we would obtain in a 3D crystal, we see that the problem arises from the missing 
density-of-states factor of $Q$ in the numerator and, most important, the additional factor of $Q$
in the denominator, resulting from the 2D nature of the system: The `spring constant' that pulls displaced ions back onto the plane
decreases with $Q$ at long wavelength, since the force is due only to the projection along the $z$ direction
of the `pull' of adjacent ions, also on the plane. This forces a `soft' parabolic dispersion for the ZA phonons. 

This divergence translates directly into divergences in the calculation of the electron/ZA-phonon scattering rate. We consider here the
specific case of silicene, measuring electron energies from the energy of the symmetry point K. We also consider the ideal case of a Dirac
dispersion, keeping in mind that oxidation\cite{Yi_2014}, dopants\cite{Ni_2014}, still controversial substrate-related effects\cite{Wang_2013},
the application of vertical electric fields\cite{Drummond_2012}, or the spin-orbit interaction\cite{Liu_2011} may open a gap, albeit only of the order of 1 meV in most cases. To anticipate the discussion of the 
following section, we assume that ZA phonons with wavelength longer than $\lambda_{0}$ 
are damped and/or stiffened with a dispersion of the form $bQ^{2}$ for $Q \ge Q_{0}$ and $bQ_{0}^{2-\alpha}Q^{\alpha}$ for $Q < Q_{0}$,
with some exponent $\alpha < 2$. 
We shall consider explicitly the two cases of a linear dispersion, $\alpha =1$, as suggested by Ong and Pop\cite{Ong_2011} for supported crystals whose 
ZA-dispersion is stiffened by coupling with substrate Rayleigh modes, and $\alpha =3/2$, as suggested by 
Mariani and von Oppen\cite{Mariani_2008,Mariani_2010} when accounting for the in-plane/flexural modes anharmonic coupling we have already mentioned.  
This is just one example of the general stiffened dispersion of the form $\sim Q^{\alpha}$ arising from this coupling.  
We shall discuss below in more detail the reasons behind this choice and the possible origin of this cutoff.
 
As a result of these assumptions, 
electrons with wavevector ${\bm K}$ can scatter by emitting or absorbing a `parabolic' ZA phonon only to a final state ${\bm K}'$ 
such that the scattering angle $\phi$ between the initial and final states satisfies:
\begin{equation}
K^{2} + K'^{2} - 2KK' \cos \phi \ge Q_{0}^{2} \ = (2 \pi/\lambda_{0})^{2} \ .
\label{eq:phi0}
\end{equation}
This results in a minimum scattering angle $\phi_{0} = \cos^{-1} [1 - E_{0}^{2}/(2E^{2})]$. 
On the contrary, for a smaller angle (and so, smaller electron energy)
the acoustic flexural mode involved in the process is a ZA phonon with a sub-parabolic dispersion $\sim Q^{\alpha}$.
Therefore, electrons with kinetic energy smaller than the critical energy $E_{0}/2 = \hbar \upsilon_{\rm F} (\pi/\lambda_{0})$ can only interact with
stiffened, sub-parabolic ZA phonons, while above this energy scattering can occur with both parabolic ($\phi > \phi_{0}$) and sub-parabolic
($\phi \le \phi_{0}$) ZA modes. 

\begin{figure}[tb]
\centering
\includegraphics[width=8.0cm]{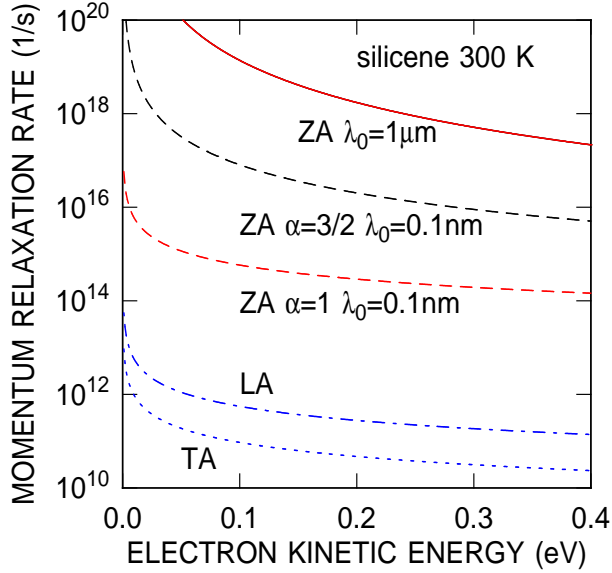}
\caption{Calculated electron momentum relaxation rates in silicene at 300~K. The momentum relaxation rate due to scattering with ZA phonons has been
         obtained assuming `stiffened' ZA-phonons ($\omega^{\rm (ZA)}_{Q} \sim Q^{\alpha}$, with $\alpha$ =1 and 3/2) for wavelengths longer than the cutoff
         $\lambda_{0}$ = 1 $\mu$m (essentially without any cutoff, yielding results largely independent of $\alpha$) and 0.1 nm, as indicated. 
         The large value of the relaxation rate obviously indicates a failure of perturbation theory
         resulting from the diverging thermal population of ZA phonons.} 
\label{fig:rates}
\end{figure}

The calculation of the scattering and momentum relaxation rates
obviously requires information about the dependence of the `deformation potential' $DK$ on the initial and final state,
${\bm K}$ and ${\bm K}'$. In order to avoid the complication of calculating the electron-phonon matrix element throughout the entire Brillouin zone,
here and in the following we choose the initial state ${\bm K}$ along particular crystallographic directions obtaining $DK$ from DFT calculations
and/or using a tight-binding model, as described above. The
choice of a particular direction should not affect the mobility in a significant way and should not distract from our main message.  
For crystals with a Dirac-like dispersion at the symmetry point K, the momentum relaxation rate expressed as a function of electron energy $E$ 
for ${\bm K}$ along the K-$\Gamma$ symmetry line and calculated using Fermi's Golden Rule and the elastic, high-temperature equipartition 
approximation takes the form:
\begin{align}
& \frac{1}{\tau_{\rm ZA}(E)} \ = \frac{2 \pi}{\hbar} \int \ \frac{{\rm d}{\bm Q}}{(2 \pi)^{2}} \ 
                     | \langle {\bm K}'+{\bm Q} | \widehat {H}_{\rm ep} | {\bm K} \rangle |^{2} \nonumber \\
& (1 - \cos \phi ) \  \delta(E_{\bm K}-E_{{\bm K}'} \pm \hbar \omega_{\bm Q}) \ ,
\label{eq:GoldenRule}
\end{align}
where $\phi$ is the scattering angle and ${\bm Q} = {\bm K} - {\bm K}'$. In this expression we have approximated the
term $1 - \upsilon({\bm K}')/\upsilon({\bm K})$ with $1 - \cos \phi$, thanks to the assumption of elastic and isotropic rate. 
Here $\upsilon({\bm K})$ is the group velocity of an electron with wavevector ${\bm K}$. 
This expression for the momentum relaxation rate does not diverge, 
since the singularity we would encounter in calculating the {\it scattering} rate would occur at 
long wavelengths and no energy or momentum are transfered in the limit $Q \rightarrow 0$. This is accounted by the factor $1 - \cos \phi$.
For a Dirac-like dispersion, with $DK$ given by Eq.~(\ref{eq:DKac_2}), and accounting for both
emission and absorption processes (identical in our approximation), this expression becomes:
\begin{equation}
\frac{1}{\tau_{\rm ZA,p}(E)} \ = \ 
\frac{DK_{0}^{2} k_{\rm B}T (\hbar \upsilon_{\rm F})^{2\alpha-2}} {2^{\alpha} \pi \hbar \rho \ b^{2} E^{2\alpha-1}} \
                        \ \int_{0}^{\pi} \ {\rm d} \phi \ \frac{\sin^{2}(\phi/2)}{(1-\cos \phi)^{\alpha-1}}  
\label{eq:relax_Dirac_lowE}
\end{equation}
for $E \le E_{0}/2$, and
\begin{align}
\hspace*{-0.25cm}
\frac{1}{\tau_{\rm ZA,p}(E)} \ = \  
& \frac{DK_{0}^{2} k_{\rm B}T (\hbar \upsilon_{\rm F})^{2\alpha-2}} {2^{\alpha} \hbar \pi \rho \ b^{2} E^{2\alpha-1}} \
                        \ \int_{0}^{\phi_{0}} \ {\rm d} \phi \ \frac{\sin^{2}(\phi/2)}{(1-\cos \phi)^{\alpha-1}} \nonumber \\ 
& + \frac{DK_{0}^{2} k_{\rm B}T \hbar \upsilon_{\rm F}^{2}} {4 \pi \rho \ b^{2} E^{3}} \
                        \ \int_{\phi_{0}}^{\pi} \ {\rm d} \phi \ \frac{\sin^{2}(\phi/2)}{1-\cos \phi} 
\label{eq:relax_Dirac_highE}
\end{align}
for $E > E_{0}/2$. For the particular case $\alpha$ = 3/2 we are considering, these expressions reduce to:
\begin{align}
\frac{1}{\tau_{\rm ZA,p}(x)} \ = \ 
& \frac{DK_{0}^{2} \upsilon_{\rm F}} {2 \pi \rho b^{2} Q_{0} x^{2}} 
    \left \{ \theta \left ( \frac{x_{0}}{2}-x \right )^{\vphantom{A}} \right. \nonumber \\
& + \theta \left (x-\frac{x_{0}}{2} \right ) \left [ 1 - \left (1- \frac{x_{0}^{2}}{4 x^{2}} \right )^{1/2} \right. \nonumber \\ 
& \left. \left. + \frac{\pi}{4} \frac{x_{0}}{x}  \left [ 1 - \frac{1}{\pi} \cos^{-1} \left ( 1- \frac{x_{0}^{2}}{2x^{2}} 
            \right ) \right ]^{\vphantom{A}} \ \right ] \ \right \} \ ,   
\label{eq:relax_Dirac_32}
\end{align}
having used the Heavyside step function $\theta(x)$ (=1 for $x>0$, =0 otherwise) and having
expressed all energies in thermal units ({\it i.e.}, $x_{0} = E_{0}/(k_{\rm B}T)$ and $x = E/(k_{\rm B}T))$. For a linear ZA-phonon dispersion,
$\alpha$ =1, instead: 
\begin{align}
& \frac{1}{\tau_{\rm ZA,p}(x)} \ = \ \frac{DK_{0}^{2}}{4 \hbar \rho b^{2} Q_{0}^{2} x} 
   \left \{ \theta \left ( \frac{x_{0}}{2}-x \right )^{\vphantom{A}} \right. \nonumber \\
& + \theta \left (x-\frac{x_{0}}{2} \right ) \left [ \frac{1}{\pi} \cos^{-1} \left ( 1- \frac{x_{0}^{2}}{2 x^{2}} \right ) - 
        \frac{1}{\pi} \frac{x_{0}}{x} \left ( 1- \frac{x_{0}^{2}}{4 x^{2}}  \right )^{1/2} \right. \nonumber \\ 
& \left. \left. + \frac{1}{2} \frac{x_{0}^{2}}{x^{2}}  \left [ 1 - \frac{1}{\pi} \cos^{-1} \left ( 1- \frac{x_{0}^{2}}{2x^{2}} 
             \right ) \right ]^{\vphantom{A}} \ \right ] \ \right \} 
\ ,   
\label{eq:relax_Dirac_1}
\end{align}
The momentum relaxation rates calculated using the cutoff-wavelengths $\lambda_{0}$ of 0.1 nm and 1 $\mu$m (the latter essentially in the no-cutoff limit) 
are shown in Fig.~\ref{fig:rates}.

The electron mobility limited by scattering with ZA phonons can be calculated from the Kubo-Greenwood expression:
\begin{align}
\mu_{\rm ZA} \ & = \ \frac{e}{\hbar n} \ \int \ \frac{{\rm d} {\bm K}}{(2 \pi)^{2}} \ \upsilon_{x}({\bm K}) \ \tau_{\rm ZA,p}(E) \ 
      \frac{\partial f}{\partial K_{x}} \nonumber \\
& \hspace*{-1.0cm} = \ \frac{e \upsilon_{\rm F}^{2}} {2 k_{\rm B} T \mathcal{F}_{1}(x_{\rm F})}  
          \int_{0}^{\infty} \ {\rm d}x \ x \ \tau_{\rm ZA,p}(x) \ f(x) \ [1-f(x)] \ ,
\label{eq:muZA_Dirac}
\end{align}
where $n$ is the electron density {\it per valley and spin-state},
$\mathcal{F}_{1}(x_{\rm F}) = \int_{0}^{\infty} dx \ x \ [ 1 + \exp(x-x_{\rm F})]^{-1}$ and 
$x_{\rm F} = E_{\rm F}/(k_{\rm B}T)$ is the Fermi energy in thermal units. In the non-degenerate limit (whatever 
this may mean in a gapless material) and in the no-cutoff limit, $x_{0} \rightarrow 0$, the ZA-phonon-limited electron mobility is:
\begin{equation}
\mu_{\rm ZA} \ = \ 24 \ \frac {4 e \rho \ b^{2} k_{\rm B}T } {\hbar DK_{0}^{2}} \
\label{eq:muZA_Dirac_nondegen}
\end{equation}

\begin{figure}[tb]
\centering
\includegraphics[width=8.0cm]{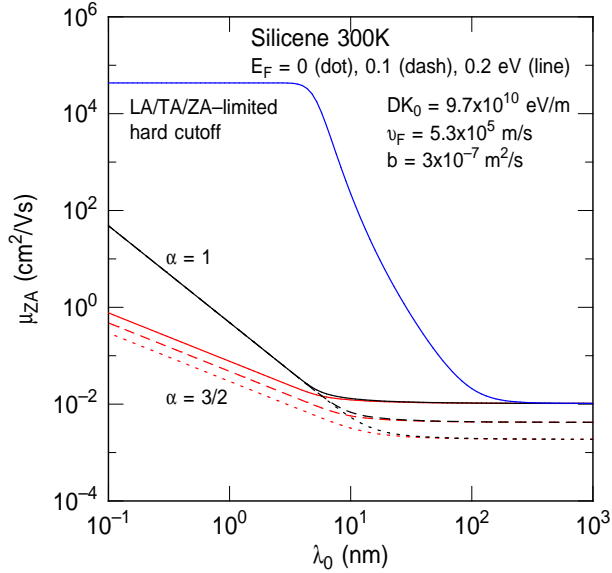}
\caption{ZA-limited electron mobility as a function of the wavelength cutoff $\lambda_{0}$ calculated for three values of the
         Fermi energy, $E_{\rm F}$ and for two different `stiffened' phonon dispersions ($\omega^{\rm (ZA)}_{Q} \sim Q^{\alpha}$ with
         $\alpha$ = 3/2 and 1) for wavelengths longer than the cutoff $\lambda_{0}$. The upper curve labeled 'hard-cutoff' is, instead,
         the electron mobility limited by in-plane (TA and LA) and flexural (ZA) acoustic modes and assuming that ZA-phonons with
         wavelengths longer than $\lambda_{0}$ are fully damped, {\em i.e.}, that electrons do not scatter with long-wavelength ZA phonons.}
\label{fig:ZA-mobility}
\end{figure}

\begin{figure}[tb]
\centering
\includegraphics[width=8.0cm]{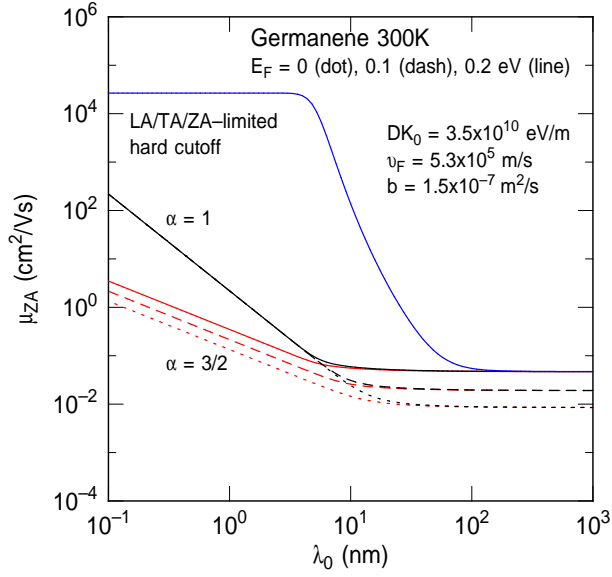}
\caption{As in Fig.~\ref{fig:ZA-mobility}, but for germanene.}
\label{fig:ZA-mobility_Ge}
\end{figure}

Using parameters valid for silicene, given in Table~\ref{tab:Parameters_D}, this results in an electron mobility of the order of
about $10^{-3}$ cm$^{2}$/Vs, essentially zero for all practical purposes. This is illustrated in Fig.~\ref{fig:ZA-mobility} in which we also show
the calculated electron mobility as a function of cutoff $\lambda_{0}$ for various values of the Fermi level (measured from the energy at the K symmetry
point). The effect of additional scattering processes (with longitudinal and transverse in-plane acoustic, as discussed below) is shown by the 
top solid line. This has been obtained by forcing a `hard cutoff' consisting in assuming that phonons with wavelength larger than
the cutoff $\lambda_{0}$ are totally damped, as it may be the case for small crystals with dimensions of the order of $\lambda_{0}$.
A similar behavior is seen for germanene, which exhibits only a marginally higher mobility, as shown in Fig.~\ref{fig:ZA-mobility_Ge}.
It should be noted that values for the mobility calculated using a cutoff $\lambda_{0}$ smaller than about 3$a_{0}$ ($\sim$ 1~nm) 
imply a stiffened ZA-phonon dispersion throughout the entire Brillouin zone, with an increasing sound velocity (in the case $\alpha$=1) that is
responsible for the increasing mobility as $\lambda_{0}$ decreases. 
Note also that the mobility increases with increasing carrier density and linearly with
temperature. This is because the momentum relaxation rate decreases at higher electron energies.

\begin{table}[tb]
\begin{center}
\caption{Parameters for `Dirac-like' materials, silicene and germanene. The angular dependence of the
         deformation potential $DK$ is given for states along the K-$\Gamma$ symmetry line. The values for the sound velocities were taken from Ref.~\onlinecite{Li_2013}.
The parameter $b$ has been estimated from the ZA phonon dispersion calculated
from first principles. Determining a more precise value for $b$ is
computationally prohibitive since a large supercell is required to
deal with small $Q$ phonons and the force constants and changes in
energy associated with flexural displacements are very small.}
\label{tab:Parameters_D}
\hspace*{-1.0cm}
\begin{tabular}{||c|cc||}
\hline
                              &           silicene          &          germanene          \\
\hline
$\rho$ (kg/m$^{3}$)           &    $7.2 \times 10^{-7}$      & $1.7 \times 10^{-6}$       \\
$c_{l}$ (m/s)                 &    $8.8 \times 10^{3}$       & $4.4 \times 10^{3}$        \\
$c_{t}$ (m/s)                 &    $5.4 \times 10^{3}$       & $2.2 \times 10^{3}$        \\
$a_{0}$ (m)                   &  $3.88 \times 10^{-10}$      & $4.03 \times 10^{-10}$     \\
$\upsilon_{\rm F}$ (m/s)      &  $5.3 \times 10^{5}$         & $5.3 \times 10^{5}$        \\
$b$ (m$^{2}$/s)               &  $3 \times 10^{-7}$          & $1.5 \times 10^{-7}$       \\
$DK_{0}$ (eV/m)             &  $9.7 \times 10^{10}$        & $3.5 \times 10^{10}$       \\
 $DK_{\rm ZA}$              &   $DK_{0} \sin \left ( \frac{\phi}{2} \right )$             & $DK_{0} \sin \left ( \frac {\phi}{2} \right )$  \\  
$DK_{\rm LA,0}$ (eV/m)      &  $2 \times 10^{9}$           & $10^{9}$                   \\
 $DK_{\rm LA}$     &   $DK_{{\rm LA},0} \sin \left ( \frac{\phi}{2} \right ) \sin(\phi)$  & $DK_{{\rm LA},0} \sin \left ( \frac{\phi}{2} \right ) \sin(\phi)$  \\
$DK_{\rm TA,0}$ (eV/m)      &  $5 \times 10^{8}$           & $1.3 \times 10^{9}$        \\
 $DK_{\rm TA}$     &   $DK_{{\rm TA},0} \sin \left (\frac{\phi}{2} \right ) \sin \left (\frac{3\phi}{2} \right )$ & $DK_{{\rm TA},0} \sin \left (\frac{\phi}{2} \right ) \sin \left ( \frac{3\phi}{2} \right )$ \\  
\hline
\end{tabular}
\end{center}
\end{table}

\begin{table}[tb]
\begin{center}
\caption{Parameters for `parabolic' materials, monolayer-HfSe$_{2}$ and iodostannanane}
\label{tab:Parameters_p}
\begin{tabular}{||c|cc||}
\hline
                              &   monolayer-HfSe$_{2}$   &          iodo-stannanane        \\
\hline
$\rho$ (kg/m$^{3}$)           &    $7.9 \times 10^{-6}$  & $3.28 \times 10^{-6}$           \\
$m^{\star}$ ($m_{0}$)         &    0.9                   & 0.2                             \\
$c_{t}$ (m/s)                 &    $5.4 \times 10^{3}$   & $2.2 \times 10^{3}$             \\
$a_{0}$ (m)                   &  $3.76 \times 10^{-10}$  & $4.9 \times 10^{-10}$           \\
$b$ (m$^{2}$/s)               &  $3 \times 10^{-7}$      & $10^{-7}$                       \\
$\Delta_{\rm ZA}$ (eV)        &         1.8              & 1.6                             \\
 $DK_{\rm ZA}$                &$\Delta_{\rm ZA} K \sin^{2} \left ( \frac{\phi}{2} \right )$&$\Delta_{\rm ZA} K \sin \left ( \frac{\phi}{2} \right ) 
                                                                                                   \sin \left ( \frac{3\phi}{2} \right )$  \\ 
                              &(at M along M-$\Gamma$)   & (at $\Gamma$ along $\Gamma$-M)  \\  
\hline
\end{tabular}
\end{center}
\end{table}

As already shown in Figs.~\ref{fig:ZA-mobility} and \ref{fig:ZA-mobility_Ge}, 
the electron mobility limited by the in-plane acoustic modes, LA and TA phonons, is very large.
In this case, for an initial wavevector ${\bm K}$ along the K-$\Gamma$ symmetry line, tight-binding and DFT calculations give the following forms 
for the deformation potential $DK$:
\begin{align}
&DK_{\rm LA} \ = \ DK_{\rm LA, 0} \sin (\phi/2) \sin(\phi)  \nonumber \\
&DK_{\rm TA} \ = \ DK_{\rm TA, 0} \sin (\phi/2) \sin(3\phi/2) \ . 
\label{LATA-DK}
\end{align}
The first of these equations shows the `usual' suppression of backwards scattering, the latter shows a threefold symmetry. Both are consistent with the
results of Ref.~\onlinecite{Gunst_2015}, 
The momentum relaxation rates associated with emission and absorption of in-plane acoustic phonons are:
\begin{equation}
\frac{1}{\tau_{\rm p, LA/TA}(E)} \ = \
     \frac{DK_{\rm LA/TA,0}^{2} k_{\rm B}T}{8 \hbar \rho c_{\rm L/T}^{2} E} \ .
\label{relax_LATA}
\end{equation}
These rates are shown by the dashed (LA) and dotted (TA) lines in Fig.~\ref{fig:rates}.
The resulting electron mobility calculated in the non-degenerate limit is:
\begin{equation}
\mu_{\rm LA/TA} \ = \
  \frac{8 e \hbar c_{\rm L/T}^{2} \rho \upsilon_{\rm F}^{2}}{(DK_{\rm LA/TA, 0})^{2} (k_{\rm B}T)^{3}} \ .
\label{mu_LATA}
\end{equation}
Using the parameters listed in Table~\ref{tab:Parameters_D} (with deformation potentials obtained from DFT
calculations), this expression yields $\mu_{\rm LA} \approx 5.06 \times 10^{4}$ cm$^{2}$/Vs 
and $\mu_{\rm TA} \approx 2.98 \times 10^{5}$ cm$^{2}$/Vs for silicene at 300~K,  
and $\mu_{\rm LA} \approx 1.34 \times 10^{5}$ cm$^{2}$/Vs,
$\mu_{\rm TA} \approx 8.82 \times 10^{4}$ cm$^{2}$/Vs for germanene. 
These are extremely high values, slightly lower than those of Ref.~\onlinecite{Shao_2013}. Other scattering processes we have intentionally
ignored here -- namely, scattering with transverse, longitudinal, and out-of-plane optical modes, TA, LO, and ZO, as well as inter-valley
processes -- also yield an electron mobility orders of magnitude larger than the ZA-limited mobility, which is the subject of our discussion.
Therefore, they do not soften our main concern: These high values show how important it is to
understand, and hopefully control, the role played by the acoustic flexural modes, for silicene -- and other 2D crystals with broken
$\sigma_{\rm h}$ symmetry -- to exhibit good electronic transport properties.

\begin{figure}[tb]
\centering
\includegraphics[width=8.0cm]{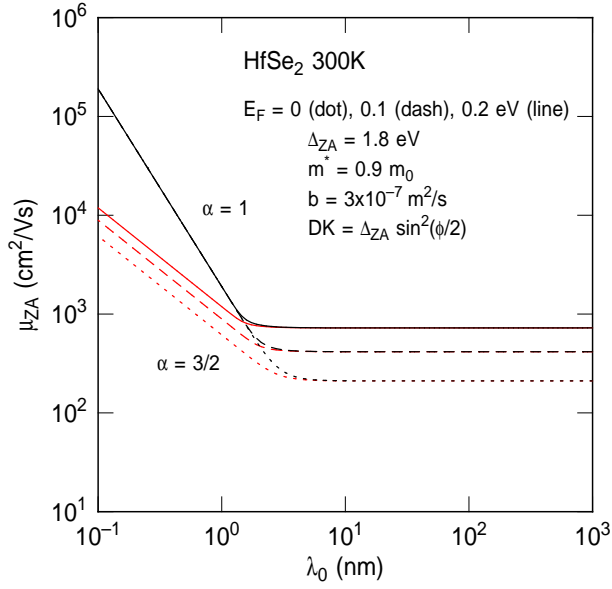}
\caption{As in Fig.~\ref{fig:ZA-mobility}, but for the `gapped and parabolic' HfSe$_{2}$.}
\label{fig:mu_parab}
\end{figure}

\begin{figure}[tb]
\centering
\includegraphics[width=8.0cm]{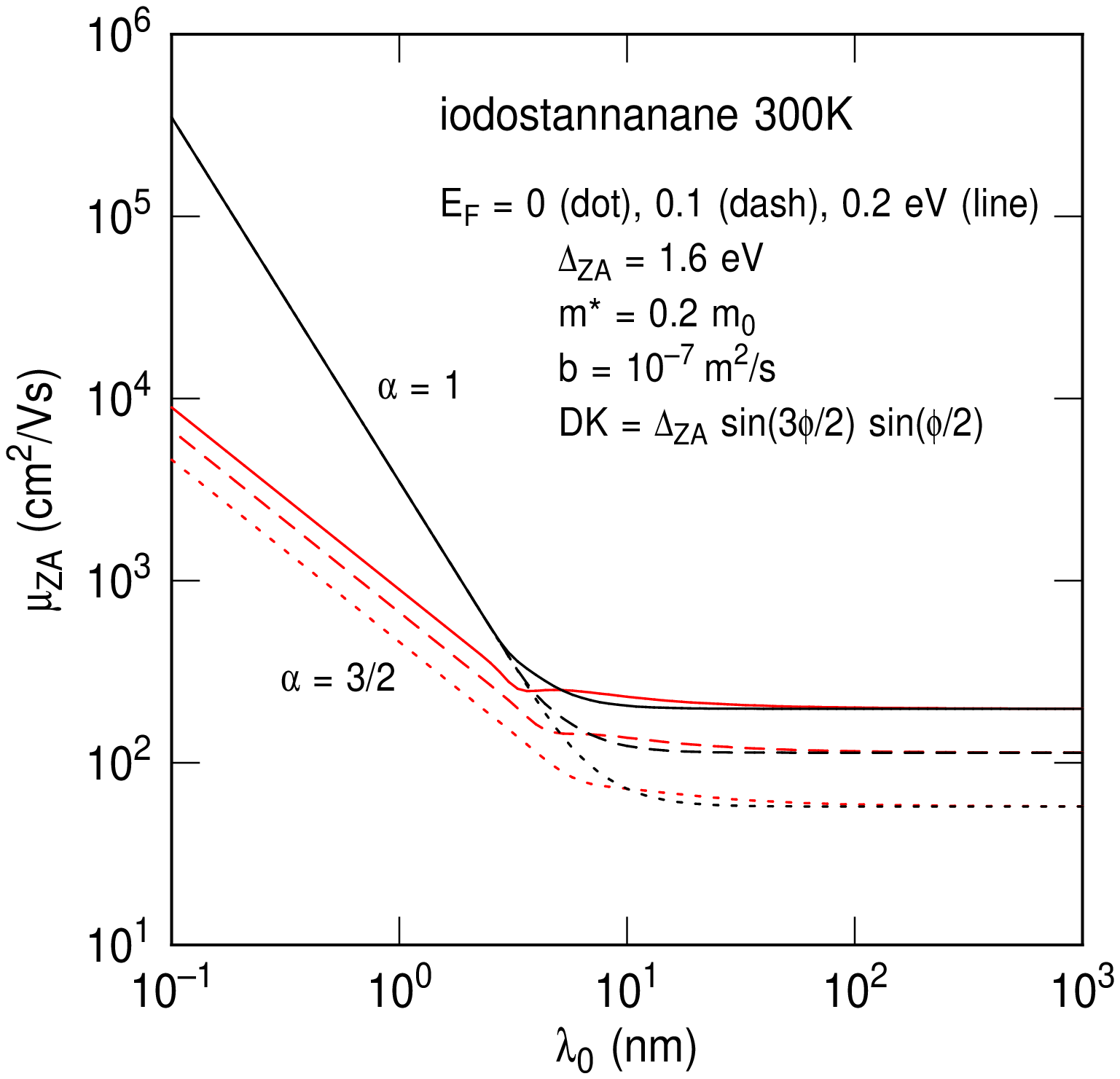}
\caption{As in Fig.~\ref{fig:mu_parab}, but for iodine-functionalized stanene, `iodostannanane'.}
\label{fig:mu_parab2}
\end{figure}

It is interesting to consider the case of `gapped' 2D materials, such as 
stanene\cite{Zhu_2015} or `stannanane' ({\it i.e.}, iodine-functionalized 
`iodostannanane'\cite{Xu_2013,Vandenberghe_2014,Vandenberghe_2014a,Suarez_2015}) and 
semiconducting TMDs with the T (or T') crystal structure. Indeed also silicene and germanene
may fall into this category when accounting for the effects of dopants, oxidation, spin-orbit interaction, substrate-related effects, or vertical fields, effects that we have already mentioned. These exhibit a parabolic electron dispersion with effective mass $m^{\ast}$ and the bottom of the conduction band is non-degenerate, so that from DFT calculations we extract a different and weaker form for the deformation potential, namely
\begin{equation}
DK_{\rm ZA} \ = \ \Delta_{\rm ZA} K \sin^{2}(\phi/2) \ ,
\label{eq:DK-M}
\end{equation}
for initial states ${\bm K}$ close to a minimum of the conduction band at the M symmetry point, such as HfSe$_{2}$, or 
\begin{equation}
DK_{\rm ZA} \ = \ \Delta_{\rm ZA} K \sin(3\phi/2) \sin(\phi/2) \ ,
\label{eq:DK-Gamma}
\end{equation}
for ${\bm K}$ close to the $\Gamma$ point, such as stannanane. In these equations $K$ is measured from the minimum of the conduction band,
M and $\Gamma$, respectively, and it is assumed to lie along the M-$\Gamma$ symmetry line.
Despite the weaker electron-ZA-phonon coupling, the parabolic dispersion 
of the ZA phonon still results in a divergent scattering rate. 
Therefore, also in this case we consider a wavelength cutoff $\lambda_{0}$ for wavelengths
below which the dispersion of the ZA phonons is stiffened to a $\sim Q^{\alpha}$ behavior. 
The resulting critical scattering angle now is $\phi_{0} = \cos^{-1} [1 - E_{0}/(2E)]$ and scattering occurs only with sub-parabolic ZA phonons  
for electron energies smaller than $E_{0}/4 = \hbar^{2}(8 m^{\ast}) (2 \pi/\lambda_{0})^{2}$. 
Accounting for both emission and absorption processes, the momentum relaxation rate can be written as:
\begin{align}
\frac{1}{\tau_{\rm ZA,p}(E)} \ & = \ \frac{ \Delta_{\rm ZA}^{2} (m^{\ast} k_{\rm B}T)^{1/2} }{4 \hbar^{2} \rho \ b^{2} \ Q_{0} \ x^{1/2}} \ \times \nonumber \\ 
& \left \{ \theta \left ( \frac{x_{0}}{4} -x \right )^{\vphantom{A}} [ A_{1}(\pi) - A_{1}(0) ] \right. \nonumber \\ 
&          + \ \theta \left ( x - \frac{x_{0}}{4} \right ) \left [ \left [ A_{1}(\phi_{0}) - A_{1}(0) \right ]^{\vphantom{A}} \right. \nonumber \\
&          \left. \left.   + \left ( \frac{x_{0}}{2x} \right )^{1/2} \left [ A_{2} (\pi)  - A_{2}(\phi_{0}) \right ]^{\vphantom{A}} 
                           \right ]^{\vphantom{A}} \right \}       
\label{eq:relax_parab_32}
\end{align}
for $\alpha$ = 3/2, and
\begin{align}
\frac{1}{\tau_{\rm ZA,p}(E)} \ & = \ \frac{ \Delta_{\rm ZA}^{2} m^{\ast} k_{\rm B}T }{2 \hbar^{3} \rho \ b^{2} \ Q_{0}^{2}} \ \times \nonumber \\ 
& \left \{ \theta \left ( \frac{x_{0}}{4} -x \right )^{\vphantom{A}} [ A_{3}(\pi) - A_{3}(0) ] \right. \nonumber \\ 
&          + \ \theta \left ( x - \frac{x_{0}}{4} \right ) \left [ \left [ A_{3}(\phi_{0}) - A_{3}(0) \right ]^{\vphantom{A}} \right. \nonumber \\
&              \left. \left. + \left ( \frac{x_{0}}{2x} \right ) \left [ A_{2} (\pi)  - A_{2}(\phi_{0}) \right ]^{\vphantom{A}} 
                      \right ]^{\vphantom{A}} \right \} \ ,      
\label{eq:relax_parab_1}
\end{align}
for $\alpha$ = 1. In these expressions the angular integrals are given by the functions:
\begin{align}
& A_{1} ( \phi ) \ = \ \frac{ \sin(\phi/2) [ \cos(3\phi/2) -9 \cos(\phi/2) ]}{6 \pi (1-\cos \phi)^{1/2}} \nonumber \\
& A_{2} ( \phi ) \ = \ \frac{ \phi - \sin\phi}{4 \pi} \nonumber \\
& A_{3} ( \phi ) \ = \ \frac{3}{8 \pi} \phi - \frac{1}{2 \pi} \sin \phi + \frac{1}{16 \pi} \sin( 2 \phi) 
\label{eq:A_Hf}
\end{align}
for HfSe$_{2}$ and
\begin{align}
\hspace*{-0.0cm}
& A_{1} ( \phi ) \ = \nonumber \\ & \frac{ \sin(\phi/2) [ -70 \cos(\phi/2) -7 \cos(3\phi/2) + 5 \cos(7\phi/2)]}{70 \pi (1-\cos \phi)^{1/2}} \nonumber \\
\hspace*{-0.0cm}
& A_{2} ( \phi ) \ = \ \frac{ 3\phi - \sin (3\phi)}{12 \pi} \nonumber \\
\hspace*{-0.0cm}
& A_{3} ( \phi ) \ = \nonumber \\ &  \frac{24 \phi = 24 \sin \phi + 6 \sin(2\phi)-8\sin(3\phi)+3\sin(4\phi)}{96\pi} 
\label{eq:A_Sn}
\end{align}
for iodostannanane.
Finally, the mobility is given by:
\begin{align}
\mu_{\rm ZA} \ & = \ 
   \frac{\pi e}{2 m^{\ast} \ln[1+\exp(x_{\rm F})]} \nonumber \\
&  \int_{0}^{\infty} dx \ x \ \tau_{\rm ZA,p}(x) \ f(x) [ 1 - f(x) ] \ .
\label{eq:muZA_parab}
\end{align}
In the non-degenerate and no wavelength-cutoff limit, Eq.~(\ref{eq:muZA_parab}) reduces to:
\begin{equation}
\mu_{\rm ZA} \ = \ 
   \frac{128 e \hbar \rho \ b^{2}}{\Delta_{\rm ZA}^{2} m^{\ast}} \ .
\label{eq:muZA_parab_nondegen}
\end{equation}
For monolayer HfSe$_{2}$, chosen as an example of a semiconducting TMD with tetragonal crystal structure, using the parameters listed in 
Table~\ref{tab:Parameters_p}, we see that the effect of the electron dispersion and of the less-singular 
matrix elements result in a mobility of about 200 cm$^{2}$/Vs in the non-degenerate, no-cutoff  limit. If not exciting, this is at least a 
reasonable value. The full dependence of the mobility on the cutoff-wavelength $\lambda_{0}$ is shown in Fig.~\ref{fig:mu_parab}.
A qualitatively similar behavior is shown in Fig.~\ref{fig:mu_parab2} for iodostannanane, that exhibits a marginally
lower mobility, about 70 cm$^{2}$/Vs in the non-degenerate, no-cutoff  limit.



\section{The wavelength cutoff $\lambda_{0}$}

The extremely disappointing electron mobility we have calculated in the previous section is the result of the Mermin-Wagner theorem, 
with the associated diverging population of long-wavelength ZA phonons. For materials with a Dirac-like electron dispersion,
it is also the result of the broken $\sigma_{\rm h}$ symmetry that yields a very strong 
interaction of electrons with the very same acoustic flexural modes.

In obtaining these results, we have considered the presence of some mechanism providing 
a long-wavelength cutoff $\lambda_{0}$. Clearly, such a cutoff is present, or
the 2D crystal itself would not exist. The problem -- for which we can offer no solution, only speculations -- is to determine 
whether the same cutoff that allows the stability of a 2D sample of finite size is also effective in yielding a technologically useful 
electron mobility.

Several cutoffs may be invoked. First, the most effective one is the total suppression of flexural modes with a wavelength longer than
$\lambda_{0}$, as shown by the top curves in Figs.~\ref{fig:ZA-mobility} and \ref{fig:ZA-mobility_Ge}. At the electron density corresponding to the
Fermi energy shown in those figures, cutoff wavelengths of about 100~nm boost the mobility to excellent values. However, 
samples used in devices at present have a finite size of several hundreds of nanometers or even tens of
micrometers. Looking at Fig.~\ref{fig:ZA-mobility}, these cutoffs are not effective in raising the
mobility to reasonably large values. Therefore, it seems unlikely that the finite size of the samples usually grown 
(1-to-10 $\mu$m for TMDs\cite{Ly_2014,Duan_2015}) or the correlation-length of the roughness and `wrinkles' usually observed experimentally
(also of the order of micrometers in most two-dimensional crystals\cite{Roldan_2015}), are sufficient to result in a technologically useful carrier mobility. Even in silicene and germanene, grown with a relatively immature technology, grain sizes exceeding several hundreds of nanometers are routinely
achieved (see, for example, Ref.~\onlinecite{Jamgotchian_2014} for silicene on Ag(111)). 

A more likely argument originates from what we have mentioned above and discussed at length in the literature in the context 
of suspended graphene: A `natural' cutoff has been shown to arise from the coupling of the 
flexural modes to the in-plane modes\cite{Castro_2010,Mariani_2008,Mariani_2010}: These are essentially strain waves that stiffen the flexural 
modes, thus stabilizing membranes\cite{Nelson_1987} and 2D crystals\cite{Gornyi_2015}.
Mariani and von Oppen\cite{Mariani_2008,Mariani_2010} have shown that this coupling changes the parabolic dispersion of the ZA phonons, yielding
a non-singular behavior $\omega^{\rm (ZA)}_{\bf Q} \sim b Q_{0}^{1/2} Q^{3/2}$ for wavevectors $Q$ much smaller that a cutoff
$Q_{0}^{2} = 3k_{\rm B}T c_{\rm t}^{2}/(2 \pi \rho_{0} b^{4})(1-c_{\rm t}^{2}/c_{\rm l}^{2})$ (expressed in our notation)\cite{noteQ0}. 
As we mentioned above,
in $\sigma_{\rm h}$-symmetric crystals, this coupling stabilizes the structure\cite{Gornyi_2015} and reduces significantly the strength of the 
electron/two-phonon interactions at the Dirac point by eliminating the singularity\cite{Gornyi_2012}. 
This depends crucially on several factors: First, the vanishing of the deformation potential at small $Q$ in these mirror-symmetric materials
reduces the severity of the long-wavelength singularity. As we have already remarked, in non-$\sigma_{\rm h}$-symmetric materials with a 
Dirac electron dispersion, the deformation potential diverges as the initial and final states approach the K point. Therefore, although the singularity 
is weakened by this `stiffening', it still persists in non-symmetric crystals. Not so in a symmetric material like graphene, for example.
Second, even ignoring `cutoffs' of any type, the strong effect of this stiffening in, say, graphene, depends on the fact that a renormalization 
of the dispersion of the acoustic flexural modes (such as $Q^{2-\eta/2}$ instead of $Q^{2}$, as in Ref.~\onlinecite{Gornyi_2012}, or even 
$Q^{2-\eta}$, as in Ref.~\onlinecite{Gornyi_2015}, with $\eta \sim$ 0.7-0.8) regularizes much 
more effectively weaker two-phonon processes, the only ones allowed in graphene, than one-phonon processes, allowed in the 
non-$\sigma_{\rm h}$-symmetric materials of interest here.    
Finally, in `heavier' materials, such as silicene and especially germanene, the in-plane modes are softer than in graphene. Therefore, the effect of the
anharmonic coupling is smaller than in graphene. For example, because of the smaller sound velocities, this coupling will result in an anharmonic cutoff 
$\lambda_{0}$ of about 3 nm and 2 nm in silicene and germanene respectively. This is still not enough to boost the mobility above 0.1 cm/Vs, as shown by
the curves labeled `$\alpha$ = 3/2' in Figs.~\ref{fig:ZA-mobility} and \ref{fig:ZA-mobility_Ge}.
On a related note, tensile strain in supported layers should not be expected to provide any stronger cutoff: 
van der Waals interactions are likely to be too weak, whereas chemical bonding with the supporting substrate would alter the electronic structure of 
the layer so much as to yield a totally different structure.

Another plausible argument is based on the observation that 
in very small samples the material may exhibit a band gap and, so, a parabolic electron dispersion. This may result from  
dopants, a strong spin-orbit interaction, or vertical electric fields, for example. However, even in this case one would not expect a satisfactory carrier mobility, as shown in Fig.~\ref{fig:ZA-mobility}. 

Perhaps the most important observation relies on the fact that
most 2D samples used to fabricate field-effect transistors, such as the silicene FET of Ref.~\onlinecite{Tao_2015},   
are `clamped' by a supporting substrate and a gate insulator. This may result in a damping of long-wavelength ZA phonons or even in 
a significant stiffening of their dispersion, as the crystal would not be free to exhibit the same strong out-of-plane oscillations
as in the free-standing case. Effects of this type have been studied by Ong and Pop using a continuum model\cite{Ong_2011}. 
They have shown that, in a single-layer graphene supported by SiO$_{2}$, if one could somehow strengthen the interaction between the graphene sheet and the substrate
(presumable by controlling in morphology of the interface), this would cause the ZA phonons to exhibit a linear dispersion as a result of the hybridization of the 
graphene flexural models with the Rayleigh waves of the substrate. Being interested in thermal transport, they conclude that this enhanced coupling with the
substrate would result in an enhanced velocity of the hybrid ZA-Rayleigh modes and in an enhanced 
thermal conductivity. However, obviously such a linearization of the dispersion of the acoustic flexural modes in non-$\sigma_{\rm h}$-symmetric crystals 
would have dramatic effects on electronic transport as well. Unfortunately, even a linear ZA-phonon dispersion does not seem to boost the mobility to
technologically useful values, as seen in the curves labeled `$\alpha$ = 1' in Figs.~\ref{fig:ZA-mobility} and \ref{fig:ZA-mobility_Ge}. 

Ong and Pop's arguments are suggestive but only speculative, since a linearization 
of the ZA-phonon dispersion requires a graphene-substrate coupling at least one order of magnitude stronger than what is the expected from van der Waals forces.
The possibility (or even likelihood) that the interaction with the substrate may damp the electron/ZA-phonon interaction has been mentioned also 
by Gunst {\em et al.}\cite{Gunst_2015} and has been shown by Glavin {\em et al.}\cite{Glavin_2002} to be an important effect in the different context of confined phonons in thin quantum wells. 

In the same spirit of Ong and Pop's arguments\cite{Ong_2011}, here we argue that any such `clamping' provided by the supporting substrate and/or gate
insulator sufficient to limit the amplitude of the out-of-plane ionic displacement to much less than 1\% of the lattice constant (probably requiring a coupling 
stronger than a simple van der Waals interaction) may be sufficient to boost significantly the carrier mobility. 
\begin{figure}[tb]
\centering
\includegraphics[width=8.0cm]{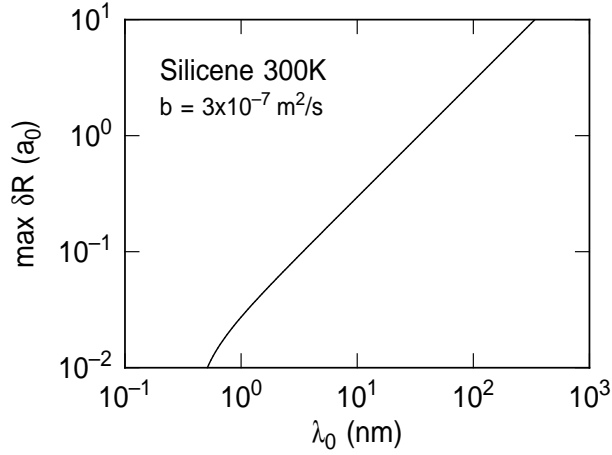}
\caption{Cutoff wavelength, $\lambda_{0}$, as a function of the maximum allowed out-of-plane displacement $\delta R$, expressed in
         units of the lattice constant $a_{0}$.}
\label{fig:cutoff}
\end{figure} 

In order to obtain a semi-quantitative idea of this last observation, we consider the thermal expectation value of the out-of-plane ionic
displacement due to a population of thermally excited ZA phonons. The squared of such ionic displacement is given by: 
\begin{align}
\left \langle \delta \widehat{R}^{\dagger} \delta \widehat{R} \right \rangle_{\rm th} & = 
    \int_{Q<Q_{0}} \frac{{\mathrm d} {\bm Q}}{(2 \pi)^{2}} \ \frac{\hbar}{2 \rho \ \omega^{\rm (ZA)}_{Q}} \  
                   \left ( 1 + 2 \left \langle N_{\bm Q}^{\rm (ZA)} \right \rangle_{\rm th} \right ) \nonumber \\
    & \approx \ \frac{k_{\rm B}T a_{0}^{2}} {16 \pi^{2} \rho \ b^{2} } \ 
           \left [ \left ( \frac{\lambda_{0}} {a_{0}} \right )^{2} \ -1 \right ] \ .
\label{eq:displacement}
\end{align}
From this, we obtain 
\begin{equation}
\lambda_{0} \ = \ a_{0} \left [ 1 + \frac{1}{\alpha^{2}} \ \frac{16 \pi^{3} \rho \ b^{2}}{k_{\rm B}T} \right ]^{1/2} \ , 
\label{eq:cutoff}
\end{equation}
a relation that expresses the cutoff $\lambda_{0}$ in terms of the maximum allowed ionic `vertical' displacement, $\alpha$, in units of the
lattice constant $a_{0}$. 
This is shown in Fig.~\ref{fig:cutoff}. Comparing this figure with Fig.~\ref{fig:ZA-mobility}, we see that for
silicene, using $a_{0}$ = 0.384~nm, an electron mobility of about 100 cm$^{2}$/Vs (as reported in 
Ref.~\onlinecite{Tao_2015}), is obtained only demanding a maximum out-of-plane displacement much smaller than 1\% of the lattice constant.
This seems a very small displacement, requiring extremely stiff substrate and gate-insulator materials. However, it is comparable to the 
thermal ionic displacement in bulk Si at 300~K. Note that also the in-plane TA and LA phonons cause a diverging ionic displacement, although 
in this case the divergence is less severe (logarithmic, as can be seen by inserting a linear phonon dispersion in
Eq.~(\ref{eq:displacement})), and a more relaxed cutoff of about 200~nm is sufficient to restrict it to 1-3\% of 
the lattice constant. Finally, note that the effect of the spin-orbit interaction in weakening the coupling of the electron/ZA-phonon by 
lifting the degeneracy at the K symmetry point may be significant in germanene, less so in silicene.

Ironically, an extremely low electron mobility is advantageous in applications exploiting the ballistic electron conduction in the edge states
of nanoribbons of 2D topological insulators, such as iodostannanane\cite{Vandenberghe_2014,Vandenberghe_2014a,Suarez_2015}. 
In these proposed devices, the on-state current is obtained at carrier densities (and Fermi levels) small enough to suppress scattering between the edge states in intra-edge which back-scattering 
is topologically-protected. In the off state, the high Fermi energy enhances the overlap between wavefunctions in opposite edges, thus boosting the
inter-edge matrix elements and the associated strong inter-edge back scattering reducing the current by 3-to-4 orders of magnitude. Therefore, the
current due to transport of charge carriers in the bulk acts as a leakage and a very low ZA-phonon-limited mobility (given by 
Eq.~(\ref{eq:muZA_parab}) when accounting for a small band-gap opening caused by the spin-orbit interactions) reduces dramatically this 
undesired off-state leakage.   

Finally, the extremely strong coupling between electrons and ZA phonons in the absence of any cutoff hints at a failure of perturbation theory.
A correct treatment of the problem would require the renormalization of electron dispersion, accounting for the formation 
of acoustic polarons. However, the strength of the coupling makes it very hard to perform such a calculation 
beyond perturbation theory. We expand these considerations in Appendix~\ref{sec:Renormalization}.

\section{Conclusions}

We have shown that ideal, free-standing, infinite 2D crystals lacking $\sigma_{\rm h}$ symmetry exhibit a very poor electron mobility, because
of the large number of thermally excited flexural (ZA) acoustic phonons. This is directly related to the Mermin-Wagner-Hohenberg-Coleman
theorem. Furthermore, 2D crystals without $\sigma_{\rm h}$ symmetry that also exhibit a Dirac-like electron dispersion at the symmetry point K suffer
from an enhanced electron/ZA-phonon coupling, as a consequence of the degeneracy of the bands at K. Differently from what is known for
$\sigma_{\rm h}$-symmetric crystals like graphene, we have speculated that, if a
long-wavelength cutoff is present (finite grain size, wrinkle correlation length), or if an anharmonic stiffening of the acoustic flexural modes arises
so the Mermin-Wagner theorem is circumvented and the crystals is stable, then this cutoff-wavelength is likely to be too large 
(or the stiffening too weak) to result in a significantly larger electron mobility. Other cutoffs -- such as clamping of the 2D layer by a supporting 
substrate and/or gate-insulator or phonon stiffening by the clamping layers --  must be present in order to obtain a satisfactory carrier mobility.

As we stated in the introduction, our goal is to point out the existence of this 'mobility/ZA-phonons' problem. 
In the same spirit of the Mermin-Wagner theorem, we have shown how this problem arises, but we cannot provide any conclusive answer to the
question of how to solve the problem, or even whether solutions are naturally provided by the non-ideality of the 2D crystals we are 
interested in.

\acknowledgments

This work has been supported by a Nanoelectronics Research Initiative/South West Academy of Nanoelectronics (NRI/SWAN) grant. The authors also thank Robert M. Wallace for constructive discussions and for having critically read the manuscript.

\appendix
\section{Renormalization and electrons and ZA-phonons}
\label{sec:Renormalization}
As stated in the main text, the strong coupling between electrons and ZA-phonons in 2D crystals with broken
$\sigma_{\rm h}$ symmetry renders perturbation theory invalid. Since electrons and ZA-phonons are so strongly coupled, we should
consider the self-consistent renormalization of the electron and ZA-phonon
dispersions, $\Delta_{\rm c,v}({\bm K})$ and $\delta_{\bm Q}$ (both in principle complex quantities). A subset of diagrams that can be summed
to all order in perturbation theory (diagrams containing only successive, non overlapping virtual emissions/absorptions of phonons, similar
to the one-Fermion loops in the random phase approximation) would lead to a system of coupled Dyson equations for these acoustic polarons:
\begin{align}
&\Delta_{\rm c,v}({\bm K}) \ = \ \lim_{s \rightarrow 0} \ \sum_{ {\bm Q}, {n={\rm c,v}}, \pm }  
 | \langle {\bm K} \mp {\bm Q}, n | \widehat{H}_{\rm ep} | {\bm K}, c \rangle |^{2} / \nonumber \\
 &  \left [ E_{\rm c}({\bm K}) + \Delta_{\rm c}({\bm K}) - E_{n}({\bm K} \mp {\bm Q}) - \right.  \nonumber \\
 & \hspace*{1.0cm} \left. \Delta_{n}({\bm K} \mp {\bm Q}) \pm \hbar \omega^{\rm (ZA)}_{\bm Q} \pm \hbar \delta_{\bm Q} + {\rm i} s \right ] \ ,
\label{eq:renorm_5}
\end{align}
\begin{align}
&\delta_{\bm Q} \ = \ \lim_{s \rightarrow 0} \ \sum_{ {\bm K}, {n={\rm c,v}} }
 | \langle {\bm K}+{\bm Q}, n' | \widehat{H}_{\rm ep} | {\bm K}, n \rangle |^{2} / \nonumber \\
 & \left [ E_{n}({\bm K}) + \Delta_{n}({\bm K}) - E_{n}({\bm K}+{\bm Q}) - \right. \nonumber \\
 & \hspace*{1.0cm} \left. \Delta_{n}({\bm K}+{\bm Q}) + \hbar \omega^{\rm (ZA)}_{\bm Q} + \hbar \delta_{\bm Q} + {\rm i} s \right ] \ .  
\label{eq:renorm_6}
\end{align}
The real part of these expressions (the principal part of the integrals) gives the renormalization of the dispersion, the imaginary part
gives the broadening (inverse lifetime, $\hbar/\tau$) of the quasiparticles. 

Notwithstanding the difficulty of solving this problem, we would have to consider the role played by others diagrams and anharmonic effects (such as
coupling of in-plane modes to flexural modes). More important, the ground
state of the system me not be reachable via perturbation theory, as in the Bardeen-Cooper-Schrieffer superconducting state. Therefore, at present
we must leave unanswered our main questions of whether or not Nature provides a spontaneous long-wavelength cutoff.

\end{document}